\documentclass[usenatbib]{mn2e}
\usepackage{graphicx}
\usepackage{epstopdf}
\usepackage{natbib}
\usepackage{verbatim}
\usepackage{amssymb,amsmath}
\usepackage{ulem}
\usepackage{color}
\usepackage[bookmarks,bookmarksnumbered,colorlinks=true, citecolor=blue, linkcolor=black]{hyperref}
\usepackage[mathscr]{eucal}

\usepackage[usenames,dvipsnames]{xcolor}

\pdfoutput=1

\renewcommand{\thefootnote}{\fnsymbol{footnote}}

\newcommand{\msun}{{\,\rm M_\odot}}

\newcommand\apj{ApJ}
\newcommand\apjl{ApJ}
\newcommand\aap{A\&A}
\newcommand\mnras{MNRAS}
\newcommand\nat{Nature}

\title[Galaxy Comoving Number Density Evolution]{Forward and backward galaxy evolution in comoving number density space}

\author[P. Torrey et al.]
       {\parbox{18cm}{Paul~Torrey$^{1,2}$\footnotemark[1], Sarah Wellons$^{3}$,  Chung-Pei Ma$^{4}$, Philip F. Hopkins$^{2}$, and  Mark Vogelsberger$^{1}$
       }\vspace{0.3cm}\\ 
         $^1 $ MIT Kavli Institute for Astrophysics \& Space Research, Cambridge, MA, 02139, USA\\
         $^2 $ TAPIR, Mailcode 350-17, California Institute of Technology, Pasadena, CA 91125, USA \\
         $^3$ Harvard-Smithsonian Center for Astrophysics, 60 Garden Street, Cambridge, MA, 02138, USA\\ 
         $^4$ Astronomy Department, University of California at Berkeley, Berkeley, CA 94720, USA\\
         }

\begin{document}

\maketitle

\begin{abstract}
Galaxy comoving number density is commonly used to forge progenitor/descendant links between observed galaxy populations at different epochs.
However, this method breaks down in the presence of galaxy mergers, or when galaxies experience stochastic growth rates.
We present a simple analytic framework to treat the physical processes that drive the evolution and diffusion of galaxies within comoving number density space.
The evolution in mass rank order of a galaxy population with time is influenced by the galaxy coagulation rate and galaxy ``mass rank scatter'' rate.
We quantify the relative contribution of these two effects to the mass rank order evolution.
We show that galaxy coagulation is dominant at lower redshifts and stellar masses, while scattered growth rates dominate the mass rank evolution at higher redshifts and stellar masses.
For a galaxy population at $10^{10} \msun$, coagulation has been the dominant effect since $z=2.2$, but a galaxy population at $10^{11} \msun$ was dominated by mass rank scatter until $z=0.6$.
We show that although the forward and backward median number density evolution tracks are asymmetric, the backward median number density evolution can be obtained by convolving the descendant distribution function with progenitor relative abundances.
We tabulate fits for the median number density evolution and scatter which can be applied to improve the way galaxy populations are linked in multi-epoch observational datasets.
\end{abstract}


\renewcommand{\thefootnote}{\fnsymbol{footnote}}
\footnotetext[1]{E-mail: ptorrey@mit.edu}

\section{Introduction}

Identifying links between progenitor and descendant galaxy populations to empirically infer galaxy evolution tracks is notoriously difficult.
Progenitor/descendant links have been forged previously by linking galaxy populations at a constant luminosity~\citep[e.g.,][]{Wake2006}, constant mass, or by isolating specific galaxy populations such as brightest cluster galaxies~\citep{Lidman2012, Lin2013, Shankar2015}.
Simple linking methods such as these become inaccurate as the galaxy population evolves with time.
This inaccuracy results in biased conclusions about the size, morphology, star formation rate, quenched fraction, etc. evolution of galaxies~\citep[e.g.,][]{vanDokkum1996, vanDokkum2001, Saglia2010}.

A more physically motivated linking method is to forge progenitor/descendant links at a fixed comoving number density based on the cumulative stellar mass function~\citep{vanDokkum2010, Bezanson2011, Brammer2011,  Papovich2011,Patel2013, Ownsworth2014, Papovich2015}.
In contrast to, e.g., fixed mass linking, the underlying assumption is that the most massive galaxies at some redshift evolve into the most massive systems at some other redshift.
Forging progenitor/descendant links at a constant comoving number density can accommodate evolution in the mass of the galaxy population, and is easily performed for any dataset where a galaxy stellar mass function is available.
Comoving number density analysis can lead to predictions about the mass~\citep{vanDokkum2010, Marchesini2014}, star formation rate~\citep{Ownsworth2014}, size~\citep{Ownsworth2014}, gas fraction~\citep{Conselice2013}, or morphology evolution of a galaxy population which would would not have been possible under more simplistic linking assumptions.

The key feature of comoving number density analysis is that galaxy mass rankings (and hence $N(>M)$, $N(>\sigma)$, or other parameters) are less prone to changes than the galaxy properties itself.
That is, galaxy masses or other properties of galaxies can evolve significantly, while assigned number density remains reasonably static among a galaxy population {\it as long as the mass rank order among a galaxy population is preserved}.
This makes constant comoving number density a better metric or linking progenitor/descendant galaxy populations together at different observational epochs compared to the underlying physical properties themselves.

Although comoving number density analysis provides a good first order approximation to forge progenitor/descendant links~\citep{Leja2013, Torrey2015b},
galaxies do not exactly move along constant comoving number density evolution tracks.
The galaxy rank order assumption required for constant comoving number density linking to work is broken by: 1) galaxy mergers and 2) stochastic growth rates.
Galaxy mergers change the total number density of galaxies~\citep[e.g.,][]{Ownsworth2014}.
Galaxies of low mass naturally ``move up" in mass rank and change their assigned number density when higher mass galaxies coagulate during merger events.
Stochasticity in growth rates (including star formation rates as well as ex-situ growth rates from mergers) introduces an element of randomness which violates the assumption that galaxies maintain their relative mass rank order.
Stochastic growth rates can lead to a change in the median mass rank of a tracked galaxy population.

Both galaxy merger rates and stochastic growth rates are naturally handled in numerical cosmological simulations.
Galaxy comoving number density analysis has been analyzed using semi-analytic models~\citep{Leja2013, Mundy2015}, abundance matching~\citep{Behroozi2013}, and hydrodynamical simulations~\citep{Torrey2015b, Jaacks2015, Clauwens2016}.
These studies have compared the mass and velocity dispersion evolution of galaxy populations using the explicitly tracked simulated galaxy merger trees to compare against the inferred evolution from an assumed constant comoving number density.
These studies agree that constant comoving number density analysis recovers the median stellar mass evolution of a galaxy population within a factor of a few (comoving number density recovers median masses a factor of $\sim$2-3 higher than explicitly tracked galaxies at redshift $\sim3$).
However, two issues remain:  
1) The factor of $\sim$2-3 error in the median mass evolution of galaxy populations is driven by a net evolution in the median number density of evolving galaxy populations, and 
2) The significant scatter in the number density evolution tracks that initially similar galaxy populations follow.

Both of these issues can be partially addressed using cosmological simulations~\citep{Leja2013, Torrey2015b, Jaacks2015}.
In~\citet{Torrey2015b} we provided fits to the median mass and number density evolution of tracked galaxy populations.  
We showed that by substituting prescribed galaxy evolution tracks of non-constant comoving number density in place of constant comoving number density the correct median evolution can be obtained.
Non-constant comoving number density progenitor/descendant links have now been applied to observational datasets~\citep[e.g.,][]{Marchesini2014, Salmon2015, Papovich2015}. 
However, the non-constant comoving number density evolution tracks do not describe the scatter in number density evolution for tracked galaxy populations, and (relatedly) require separate fits to define the forward and backward median number density evolution.

In this paper we extend the analysis presented by in~\citet{Torrey2015b} by quantifying the scatter in the number density evolution of tracked galaxy populations and show how this scatter links the forward and backward galaxy number density evolution tracks.
The primary goals of this paper are: 
1) to explore the relationship that exists between forward and backward number density evolution rates via their intrinsic scatter 
2) to provide tabulated rates for the number density evolution and scatter that can be applied to observational galaxy selection and analysis, and
3) to consider the relative importance of galaxy coagulation and scattered/stochastic growth rates in driving number density evolution
We provide general fits to the dispersion of a galaxy population in number density space as a function of time from its initial selection.
We also present an analytic framework that relates the forward and backward evolution of galaxies in number density space based on their scatter rates.
We show that using this simple framework the asymmetry in the forward and backward number density evolution rates can be broadly captured.  
Fitting functions are provided that can be applied to observational data sets to track galaxy populations, including both the median number density evolution and scatter.
We also address the relative importance of scattered growth rates and galaxy mergers in driving galaxy number density evolution.

The structure of this paper is as follows:
In Section~\ref{sec:AA} we outline a basic formalism for tracking galaxy populations in number density space.
This includes a simple relation between the distribution of galaxies in number density space when tracked forward and backward in time.
In Section~\ref{sec:PhysicalProcesses} we break down the total/net number density evolution rate in terms of the two underlying processes: galaxy coagulation and scattered growth rates.
We quantify the relative impact of coagulation and scatter and explore the galaxy masses and redshift ranges where each effect dominates.
In Section~\ref{sec:Discussion} we discuss our results including the potential impact for the interpretation of observational data.
We conclude and summarize in Section~\ref{sec:Conclusions}.

\section{Analytic treatment of Galaxy Number Density Evolution}
\label{sec:AA}

\begin{figure*}
\centerline{\vbox{\hbox{
\includegraphics[width=2.3in]{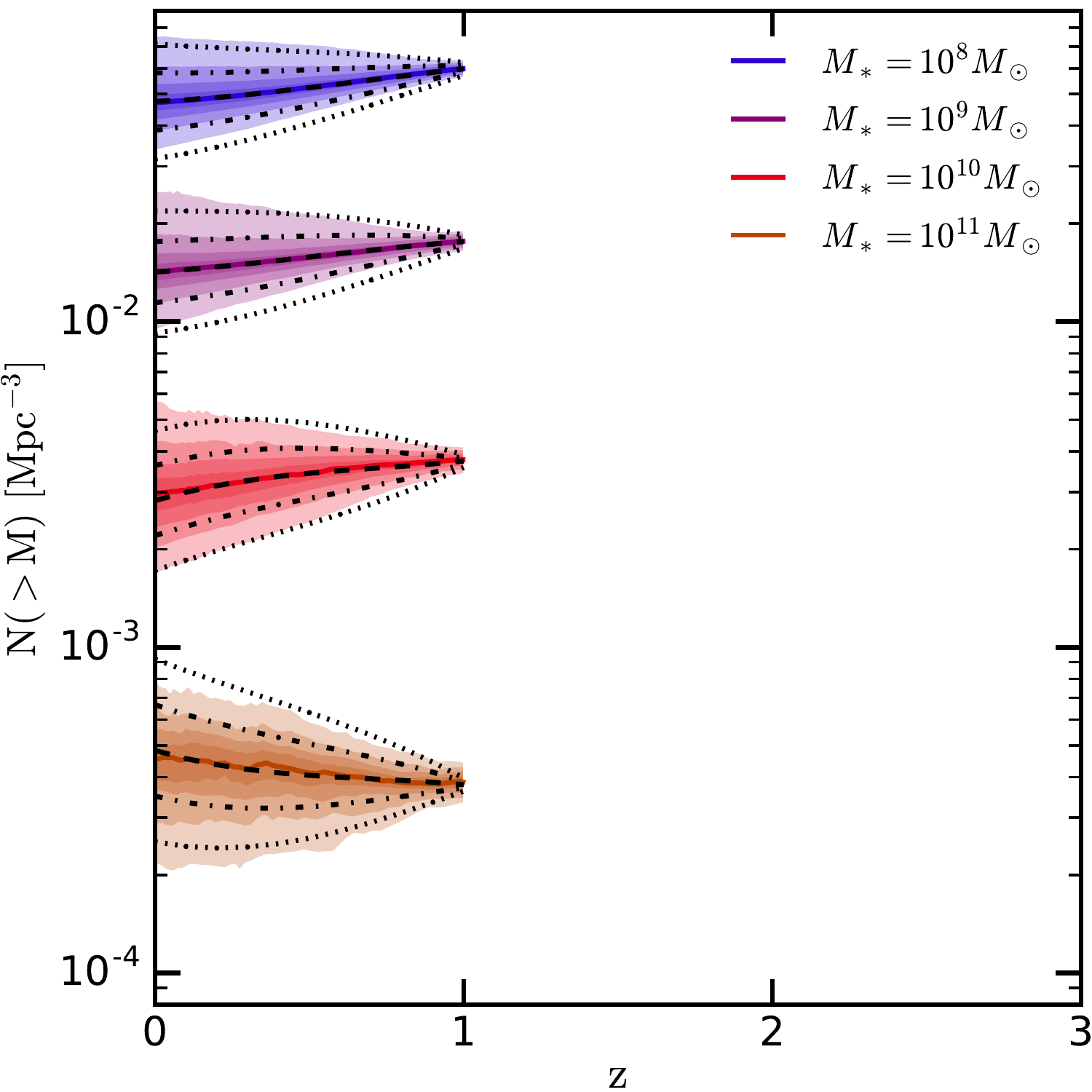}
\includegraphics[width=2.3in]{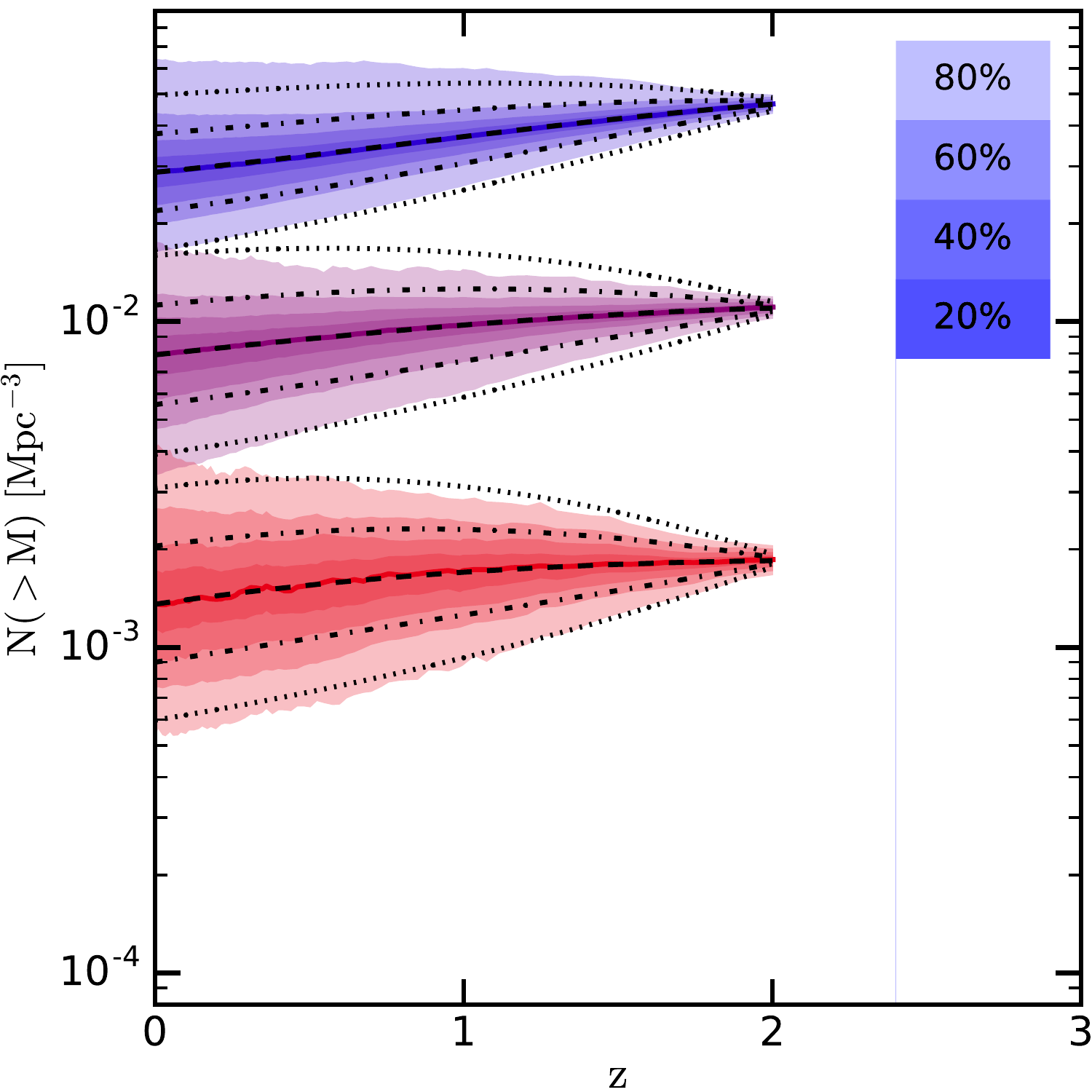}
\includegraphics[width=2.3in]{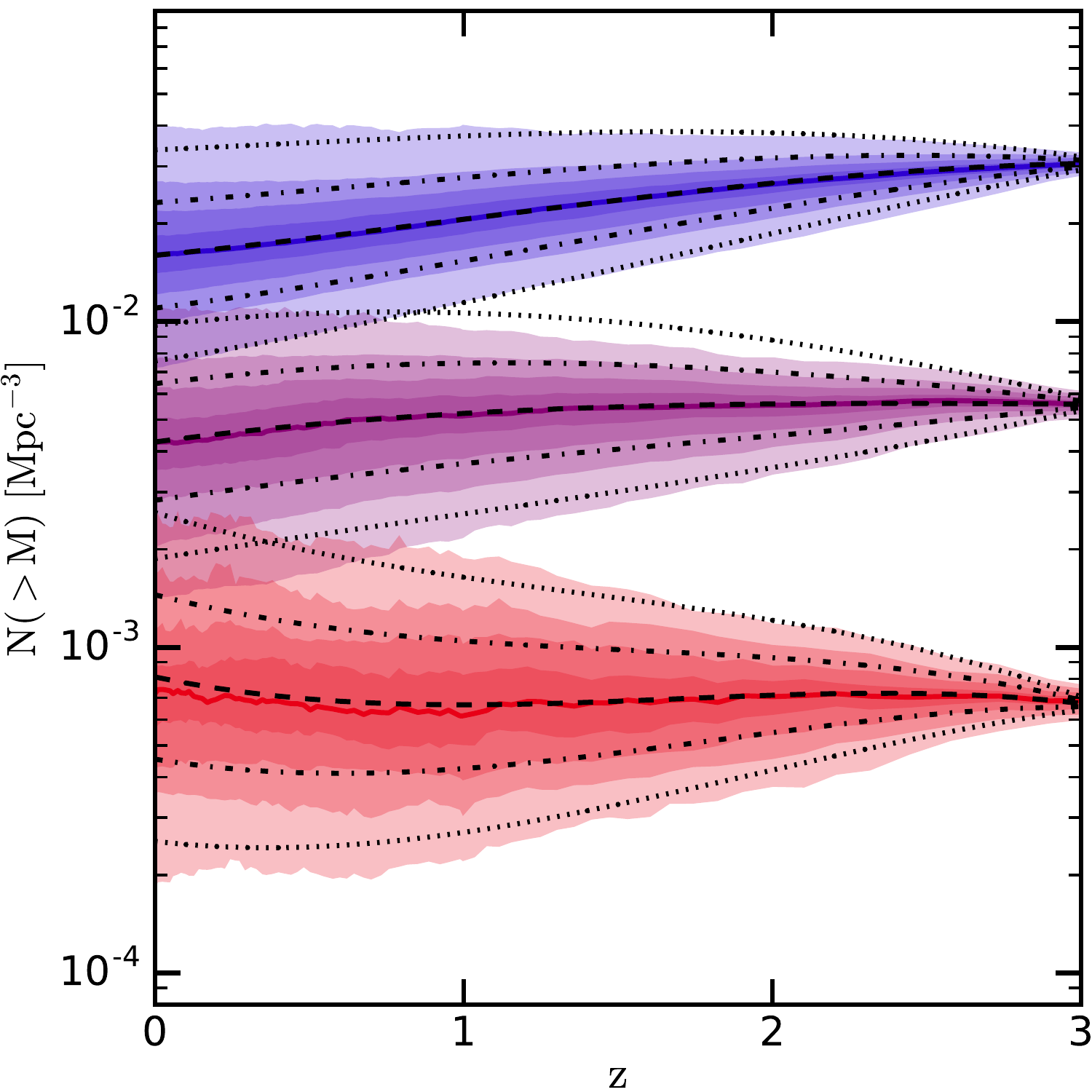}}}}
\caption{The number density evolution of galaxy populations is shown tracked in time from redshifts $z=1$, $2$, and $3$ to redshift $z=0$, from left panel to right panel respectively.
The colored shaded bands indicate the number density evolution as directly determined from the Illustris simulation -- with the shade of the color indicating the enclosed galaxy fraction as indicated in the legend.
The black dashed lines denote the best fits to the median $N$ provided in Equation~\ref{eqn:foward_num_dens_evo}.
The black dot-dashed and dotted lines denote regions of one and two $\sigma_{\rm log N}$ respectively, given in Equation~\ref{eqn:foward_scatter_evo}.
 }
\label{fig:foward_num_dens_evo}
\end{figure*}


Within a comoving volume, $V$, we rank the galaxies in the order of decreasing stellar mass and assign each galaxy a rank $R$ where $R=1,2,3,$ etc.  
A galaxy with mass $M$ -- and associated rank $R$ -- will have a cumulative number density $N = R/V$.
Mass, rank, and cumulative\footnote{
We refer to cumulative number density simply as number density throughout the paper.  
However, for clarity, all analysis and plots use cumulative number density $N(>M)$ in this paper.} number density are therefore exactly interchangeable.
However, we present our analysis in terms of number density arguments because number density is less prone to changes compared to galaxy mass.
Rank and number density can also be assigned via dark matter halo mass, velocity dispersion, or other appropriate property.
However, we have found in previous work \citep{Torrey2015b} that using stellar velocity dispersion or dark matter halo mass to assign rank produces similar number density evolution as stellar mass.  
We therefore assign rank and number density according to stellar mass throughout this paper.

We consider how a galaxy population selected within a narrow range in number density would evolve with time.
Since each of the selected galaxies from the population follows its own distinct evolution in mass ordered rank, the number density of the descendant galaxy population is best described with a distribution function.
The probability of a galaxy with initial number density $\mathcal N_0 = \mathrm{Log}(N_0)$ at redshift $z_0$ evolving to have a number density $\mathcal N_f = \mathrm{Log}(N_f)$ after time $\Delta z$ can be described as $P( \mathcal N_f | \mathcal N_0, z_0,  \Delta z ) d\mathcal N_f$.
Hereafter, we refer to $P( \mathcal N_f | \mathcal N_0, z_0,  \Delta z )$ as the descendant distribution function (DDF) which describes the distribution of number densities into which an initially homogeneous galaxy population evolves.

Not every galaxy in the initial population will necessarily survive until $z+\Delta z$; some will be consumed in galaxy mergers.
The integral of the DDF over all descendant masses therefore yields
\begin{equation}
\int _ 0 ^\infty P\left( \mathcal N_f | \mathcal N_0,  z_0, \Delta z \right) d \mathcal N_f  = f_{\mathrm{s}} \left( \mathcal N_0,  z_0, \Delta z\right)
\label{eqn:f_survival}
\end{equation}
where $ f_{\mathrm{s}}$ is the galaxy survival fraction.
The mean log number density of the descendant galaxy population is given by
\begin{align}
\left<  \mathcal N_f \right> = \frac{ \int _ 0 ^\infty \mathcal N_f P\left( \mathcal N_f | \mathcal N_0, z_0,  \Delta z \right) d \mathcal N_f}{ f_{\mathrm{s}}\left( \mathcal N_0,  z_0, \Delta z\right) }.
\label{eon:descendant_avg_rank}
\end{align}
Evaluating equation~\ref{eon:descendant_avg_rank} requires specifying a form of the DDF which we provide in Section~\ref{sec:DDF}.

\subsection{Descendant Distribution Functions}
\label{sec:DDF}

The DDF can be  determined empirically based on numerical simulations.
We find functional forms and best fits to the DDF in this section using the Illustris simulation~\citep{Vogelsberger2014a, Vogelsberger2014b, Genel2014}.
The Illustris simulation employs a model for galaxy formation~\citep{Vogelsberger2013, Torrey2014} which is able to broadly reproduce the evolution of the galaxy stellar mass function out to redshift $z=6$~\citep{Genel2014}. 

Figure~\ref{fig:foward_num_dens_evo} shows the evolution of the distribution of four galaxy populations as they are tracked in time from an initial selection redshift of $z_0=1$, $2$, and $3$ (left to right panels, respectively).
The tracked galaxy populations are selected to have stellar masses of $M_*=10^8$, $10^9$, $10^{10}$, and $10^{11} \msun$ in bins of $0.15$ dex width at their initial selection redshift.
The masses of all galaxies are tracked forward in time using the merger trees described in~\citet{RodriguezGomez2015}.
Galaxies that  are consumed in a merger event are included until they are consumed.
Masses are converted to number densities by inverting the tabulated fitting functions to the cumulative galaxy stellar mass function from~\citet{Torrey2015b}.

The DDFs presented in Figure~\ref{fig:foward_num_dens_evo} are reasonably well described by a log normal distribution
\begin{equation}
P\left( \mathcal N_f | \mathcal N_0, z_0,  \Delta z \right) = \frac{f_{\mathrm{s}}}{\sigma \sqrt{2 \pi}} \exp \left( - \frac{ \left(\mathcal N_f - \left< \mathcal N _f \right>\right)^2}{2 \sigma^2} \right)
\label{eqn:DDF}
\end{equation}
where $< \mathcal N _f >$ is the mean number density (in log space) of the descendant population, and $\sigma$ is the standard deviation.
The survival fraction, $f_{\mathrm{s}}$, mean number density $< \mathcal N _f >$, and spread, $\sigma$, are functions of the initial number density, initial selection redshift, and elapsed time.
We construct fits to $f_{\mathrm{s}}$,  $< \mathcal N _f >$, and $\sigma$ which are presented in Appendix~\ref{sec:Fits}.
Figure~\ref{fig:foward_num_dens_evo} indicates the fits to the median number density evolution track (dashed lines), the $\pm$ one-sigma fits (dot-dashed lines), and the $\pm$ two-sigma fits (dotted lines) for the tracked galaxy populations.

These fits accurately capture the median evolution and and broadly capture the scatter evolution found in the simulation.
The median number density evolution follows non-constant number density evolution tracks with evolution.
The magnitude of the change in number density varies based on initial selection redshift and initial selection mass, but ranges from roughly constant to changes of $\sim 0.3$ dex evolution out to redshift $z=3$.
The scatter grows at an approximate rate of $\sigma \propto 0.2 \Delta z$, with more detailed fits given in Appendix~\ref{sec:Fits}.

\subsection{Progenitor Number Density Distribution Functions}
\label{sec:PDF}
\begin{figure}
\centerline{\vbox{\hbox{
\includegraphics[width=3.25in]{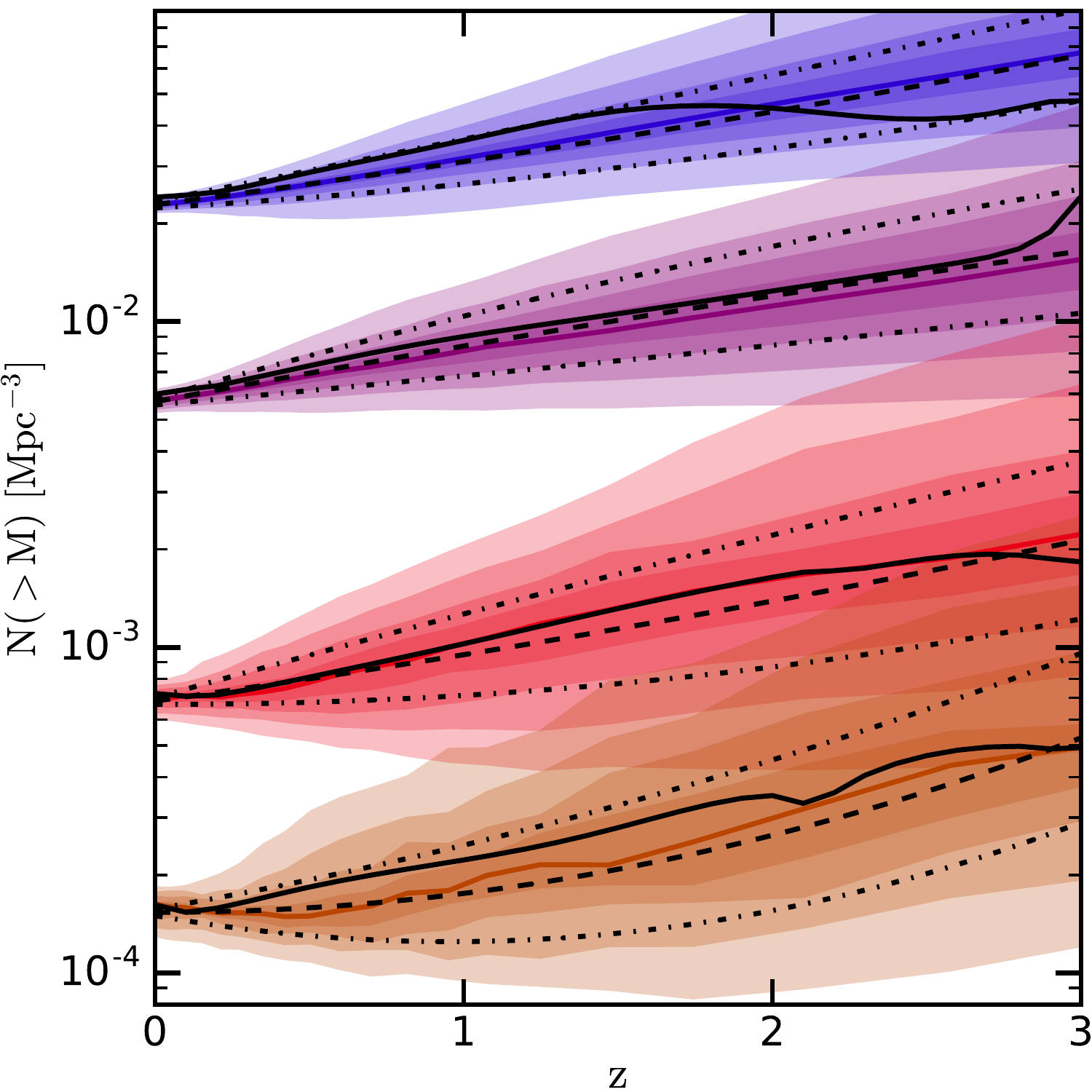}}}}
\caption{
Same as Figure~\ref{fig:foward_num_dens_evo}, but for four galaxy populations tracked backward in time.  
Solid colored lines and regions indicate the distribution of explicitly tracked galaxy populations via merger tree.
The black solid lines indicate the inferred median progenitor distribution function based on Equation~\ref{eqn:prog_avg_rank}.
The black dashed lines indicate the best fit median progenitor number density based on Equation~\ref{eqn:backward_num_dens_evo}.
The black dot-dashed lines indicate the best fit one-sigma distribution about the median progenitor number density based on Equation~\ref{eqn:backward_scatter_evo}.
The broad match between the explicitly tracked and inferred median progenitor number density out to redshift $z\sim2.5$ validates the relationship between the descendant and progenitor distribution functions presented in equation~\ref{eqn:prog_avg_rank}.
 }
\label{fig:backward_num_dens_evo}
\end{figure}

The discussion to this point has been limited to the DDF.
We can similarly consider the progenitor distribution function (PDF), which can be examined using the same numerical simulations.
Figure~\ref{fig:backward_num_dens_evo} shows the PDF for four galaxy populations selected in thin mass (number density) bins at redshift $z=0$.
Direct fits to the mean number density evolution and scatter evolution can be found in Appendix~\ref{sec:Fits}.
The direct fits are shown with black solid lines in Figure~\ref{fig:backward_num_dens_evo} which track the simulation median number density evolution very well by construction.

Examination of Figures~\ref{fig:foward_num_dens_evo} and~\ref{fig:backward_num_dens_evo} shows that the median evolution of the PDF and DDF have different slopes. 
This is a real consequence of the median number density evolution rate being different for galaxy populations that are tracked forward and backward in time~\citep[see Figure 6 of][]{Torrey2015b}.
However, the PDF and DDF can be related by considering the mean mass rank of the progenitor galaxies that will grow into galaxies with mass rank $\mathcal N_f$.  
To achieve this we need to consider the galaxies that will scatter/evolve into a particular descendant bin given the relative abundance of progenitors and the DDF.
Specifically, the mean log number density of the progenitor galaxy population that will evolve to have a number density $\mathcal N_f$ after a time $\Delta z$ is
\begin{equation}
\left<  \mathcal N_0 \right> = \frac{ \int _ 0 ^\infty \mathcal N_0  P\left( \mathcal N_f | \mathcal N_0,  z_0, \Delta z \right)  \frac{dn}{d\mathcal N_0}  d \mathcal N_0    }{   \int _ 0 ^\infty   P\left( \mathcal N_f | \mathcal N_0, z_0,  \Delta z \right)  \frac{dn}{d\mathcal N_0}  d \mathcal N_0    }
\label{eqn:prog_avg_rank1}
\end{equation}
where the integration is over all possible progenitor galaxies $d \mathcal N_0$  and $P( \mathcal N_0 | \mathcal N_f, z_0,  \Delta z )$ is  {\it the same} DDF described in the previous subsection.
The factor of $dn/d\mathcal N_0$ describes the relative abundance of the possible progenitor galaxies.
Specifically, in analogy to the galaxy stellar mass function -- where the distribution of galaxies in stellar mass is described via $\phi = dn /d \mathrm{Log} M$ -- the distribution of galaxies in number density space can be described as $dn/d\mathrm{Log}N$.
However, unlike the stellar mass function, galaxies are trivially distributed in number density space such that $dn/d\mathrm{Log}N = N = 10^{\cal N}$.
We can therefore express equation~\ref{eqn:prog_avg_rank1} as
\begin{equation}
\left<  \mathcal N_0 \right> = \frac{ \int _ 0 ^\infty \mathcal N_0  P\left( \mathcal N_f | \mathcal N_0,  z_0, \Delta z \right)  10^{\mathcal N_0}  d \mathcal N_0    }{   \int _ 0 ^\infty   P\left( \mathcal N_f | \mathcal N_0, z_0,  \Delta z \right) 10^{\mathcal N_0}  d \mathcal N_0    }.
\label{eqn:prog_avg_rank}
\end{equation}
The integrands of Equation~\ref{eqn:prog_avg_rank} are performing a convolution of (a) the probability that a galaxy with initial number density $ \mathcal N_0$ evolves into a galaxy with number density $\mathcal N_f$ with (b) the relative abundance of galaxies with initial number density $ \mathcal N_0$.
The relative abundance factor was not present in the descendant average mass rank calculation and is the cause of the offset between the forward and backward number density evolution tracking.

The physical asymmetry is driven by the relative over-abundance of fast growth tracks owing to the higher abundance of low mass galaxies.  
The absolute contribution of a progenitor bin must take into account both the likelihood that a galaxy would scatter into the desired descendant bin, and the relative abundance of the progenitor galaxy population.
Using Equations~\ref{eqn:DDF} and~\ref{eqn:prog_avg_rank} the average mass rank of the progenitor galaxy population can be inferred.
The resulting inferred number density evolution trajectories are indicated with black dashed lines in Figure~\ref{fig:backward_num_dens_evo}.
The inferred number density evolution is  within $\sim$0.05 dex out to redshift z=2.5.
Beyond redshift $z=2.5$ the fit becomes poor for the two low mass bins, owing to significant contributions to the progenitor galaxy population from outside of the fitting function validity region.
The inferred number density for the two more massive bins (red and brown) remains in broad qualitative agreement with the simulation data out to $z=2.5$.

The errors in the inferred number density evolution are driven largely by deviations from the DDF from a strict log-normal distribution.
Although the log normal DDF assumption is in broad agreement with the empirically derived DDF from the simulations, 
there are asymmetric features of the DDF which become increasingly 
prominent as the galaxy population is tracked further in redshift.  
The offsets seen in Figure~\ref{fig:backward_num_dens_evo} are small enough to give confidence that our method of linking progenitor distribution functions and descendant distribution functions is correct, but will always be less accurate than the direct fits provided in Appendix~\ref{sec:Fits}.

Finally, we note that demonstrating the time-reversibility of descendant and progenitor tracking is made possible here by the definition of a single, continuous fitting function in $\mathcal N_0$, $z_0$, and $\Delta z$ for the median and scatter of the DDF.
Such continuous fits have not been previously provided.

\section{Relating Number Density Evolution to Physical Processes}
\label{sec:PhysicalProcesses}

In this section we discuss the physical processes that drive the evolution in the number density tracks shown previously.
Specifically, we quantify the importance of galaxy coagulation and stochastic galaxy growth rates toward number density evolution and
compare the relative importance of these two mechanisms.

The average (median) number density evolution discussed at length in the previous section can be described as
\begin{equation}
\left< N (z ) \right> = N(z_0) + \int _{z_0} ^z   \left< \frac{dN}{dz} \right> dz
\end{equation}
where $ \langle dN / dz \rangle$ is the rate of change of the average number density in linear (not log) space.  
Using this form, we consider the physical processes that drive the average number density evolution.
Populations of galaxies change their median mass rank either 
1) because a galaxy moves up in mass rank (in order of decreasing mass) by one when two galaxies above its mass merge to form a single galaxy
{\bf or} 
2) by rapid or slow mass growth, such that the galaxies change their mass rank relative to their peers.
{\it These two processes provide a complete basis that can capture all of the galaxy mass assigned rank order evolution.}
Moreover, these two processes mirror the breakdown of the two fundamental assumptions of constant comoving number density analysis:  that galaxy mergers do not significantly impact the total number density of galaxies, and that galaxies preserve their mass assigned rank order in time.
In the previous section we showed how the average number density evolves for tracked galaxy populations in time.  
In this section, we discuss the relative importance that galaxy coagulation  and scattered growth rates have on comoving number density evolution.

The median number density rate of change for a galaxy population can be expressed with explicit dependence on the two number density evolution channels
\begin{equation}
\left< \frac{dN}{dz}\right> =   \left< \frac{dN_{\mathrm{c}}}{dz} \right>+ \left< \frac{dN_{s} }{dz}\right>
 \label{eqn:dn_dz}
\end{equation}
where 
$\left< dN_{\mathrm{c}}/dz \right>$ gives the rate of change of cumulative number density owing to changes in the total galaxy number density via mergers above the mass scale of interest (hereafter: the ``galaxy coagulation" rate) and 
$\left< dN_s/dz \right> $ gives number density rate of change from scattered  growth.
Equation~\ref{eqn:dn_dz} provides a rigorous and clear breakdown of the total number density evolution rate that facilitates the analysis in the subsequent subsections.
However, it does not constitute a unique breakdown of the total number density evolution rate.
For example, while galaxy coagulation is driven by galaxy mergers above a given mass scale (as described in detail below) the galaxy scatter rate also contains a contribution from ex-situ mass growth driven by galaxy mergers.
We therefore emphasize that our primary motivation for adopting the breakdown in Equation~\ref{eqn:dn_dz} is that the two term mirror the two fundamental assumptions of constant comoving number density analysis.

Galaxy coagulation drives intuitive net changes in the number density of a galaxy population by making total galaxy number density a non-conserved quantity.
It is also fairly intuitive that an individual galaxy can undergo a change in its mass ordered {\it rank} (and therefore can adjust its assigned number density) by growing much faster or slower than its peers.
It is somewhat less intuitive to predict the impact that relative galaxy growth rates have on the mass rank order evolution when averaged over a galaxy population.
However, as shown below, the contribution of scatter to the median mass rank of a galaxy population is significant when compared to the coagulation rate for a range of galaxy masses and redshifts.

\subsection{Galaxy Coagulation}
A galaxy of mass $M_i$ will necessarily change mass rank if two galaxies both with masses $M > M_i$ merge together.
When this happens, the number of galaxies with mass $M>M_i$ will decrease by exactly one, forcing all lower mass galaxies to move up in mass rank.
We therefore define the galaxy coagulation rate as the rate at which galaxies with mass rank  higher than $i$ are being swallowed by mergers (i.e. undergoing mergers with larger systems).\footnote{We briefly note that galaxy mergers also change galaxy mass, which can cause mass rank/number-density evolution.  
However, in our analytic framework, we explicitly separate changes in the total number density of galaxies (strictly driven by galaxy mergers) from mass rank order evolution driven by stochastic growth rates (which has contributions from in-situ and ex-situ/merger growth). }

\begin{figure}
\centerline{\vbox{\hbox{
\includegraphics[width=3.5in]{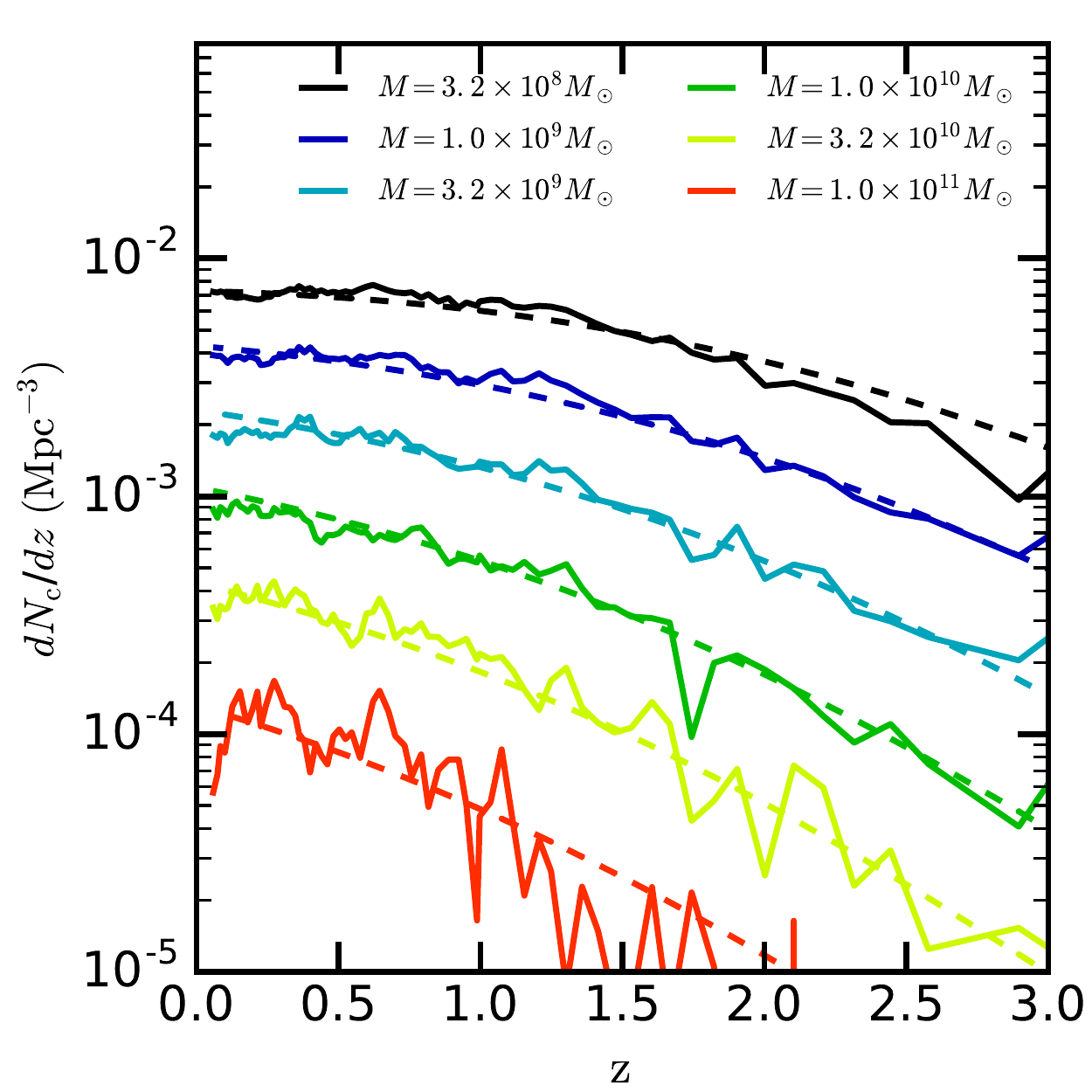}}}}
\caption{The galaxy coagulation rate as a function of time for several galaxy mass bins.  
The galaxy coagulation rate describes the rate at which a galaxy's mass rank is expected to change owing to the coagulation of galaxies with higher mass rank (higher mass).
Unlike Figures~\ref{fig:foward_num_dens_evo} and~\ref{fig:backward_num_dens_evo} where we tracked galaxy population evolution, 
here we show the galaxy coagulation rate for the same fixed $M_*$ at all redshifts, as indicated in the legend.
Dashed lines indicate the fits used in Appendix~\ref{sec:Fits} to assess the relative importance of galaxy coagulation and scattered growth rates. 
The galaxy coagulation rate can be compared against the competing galaxy scatter rate presented in Figure~\ref{fig:foward_scatter_rate}.   
}
\label{fig:destructive_merger_rate}
\end{figure}

We calculate the galaxy coagulation rate in the simulation by identifying all galaxies with initial mass $M>M_i$ which are not the main progenitor of a halo in a subsequent snapshot --  indicating that the galaxy has been consumed by a larger system in a merger event.
This number is converted to the coagulation rate using a first-order finite difference scheme with a target redshift step size of $\Delta z \approx 0.1$.  
The resulting coagulation rate is shown in Figure~\ref{fig:destructive_merger_rate}.
The coagulation rate is lower for high mass systems compared to their low mass counterparts owing simply to the lower abundance of high mass systems; the effect of galaxy coagulation rate on number density is cumulative.

The coagulation rate $dN_c/dz$ is equal to the galaxy-galaxy merger rate integrated over appropriate parameters.
Traditionally the galaxy-galaxy merger rate is tabulated as the number of mergers per unit redshift per unit mass ratio per halo, $ {\rm d} N_m / {\rm d }\xi {\rm d}z$~\citep[e.g.,][]{Fakhouri2008, RodriguezGomez2015}.
The galaxy coagulation rate can be obtained via the galaxy-galaxy merger rate by integrating over the appropriate limits
\begin{equation}
\frac{dN_{c}}{dz}(M,z)  =   \int^{\infty}_{M} dM_d \int^{1}_{M/M_p} d\xi \frac{{\rm d} N_m}{{\rm d }\xi {\rm d}z}  \frac{dn}{dM_d}
\label{eqn:red_merg_rate}
\end{equation}
where $M_p$ is the primary galaxy mass, $\xi = M_s / M_p$ is the mass ratio of the secondary to the primary, and $M_d = M_s + M_p$ is the merger descendant mass.
The additional mass function dependence in the integrand is used to convert the ``per halo" into a ``per volume".
Adopting the tabulated and parameterized merger rates of~\citet{RodriguezGomez2015}, the best-fit galaxy stellar mass functions from the appendix of~\citet{Torrey2015b}, and integrating Equation~\ref{eqn:red_merg_rate} yields an agreeable result to the direct measurement presented in Figure~\ref{fig:destructive_merger_rate}. 
Equation~\ref{eqn:red_merg_rate} reduces the three-parameter dependence of the total merger rate expression ($M_p$, $\xi$, $z$) to a two-parameter dependence on mass $M_i$ and redshift $z$.
This parameter reduction is associated with the limitations in the mass ratio that restrict mergers to only include systems which are both more massive than $M_i$.

\begin{figure*}
\centerline{\vbox{\hbox{
\includegraphics[width=3.5in]{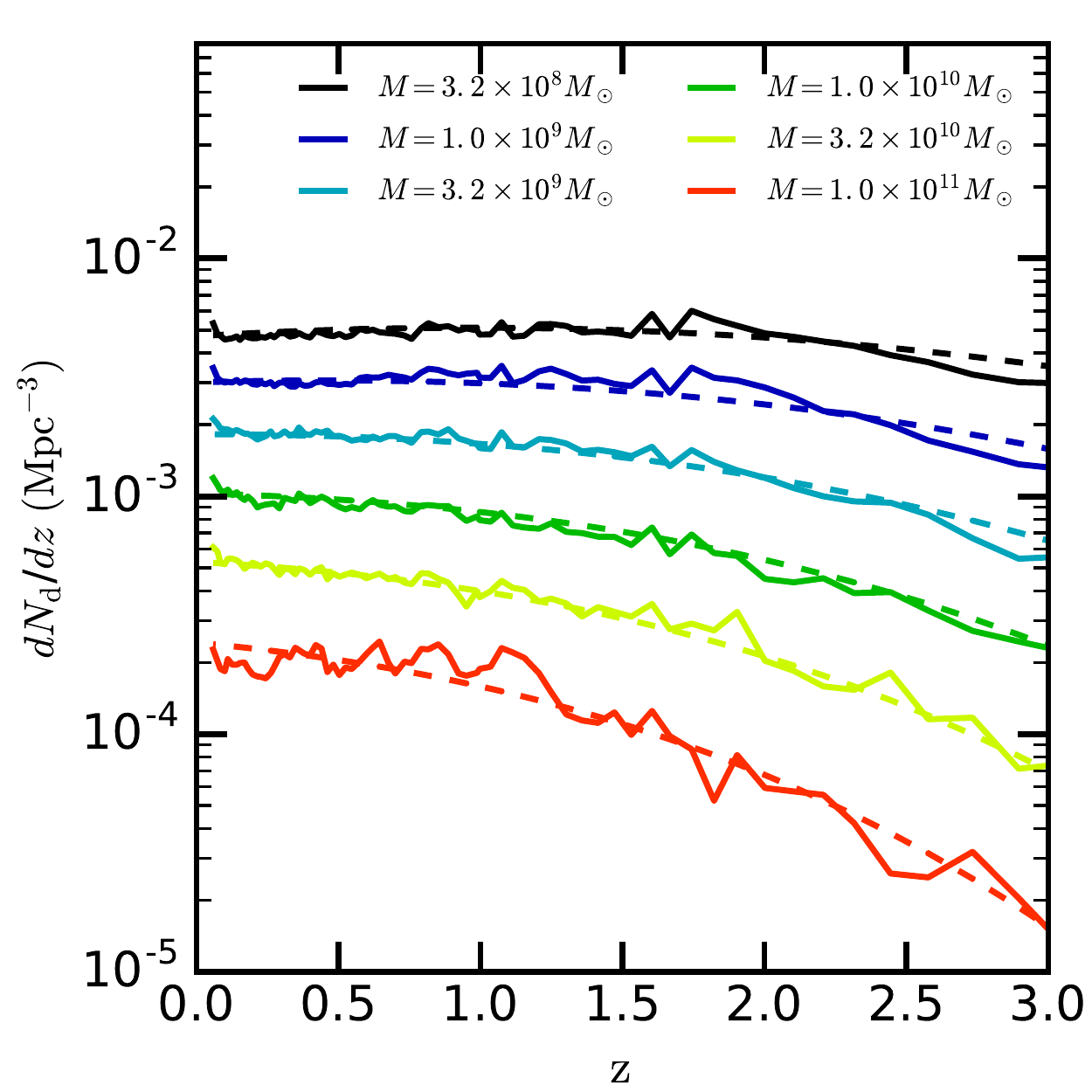}
\includegraphics[width=3.5in]{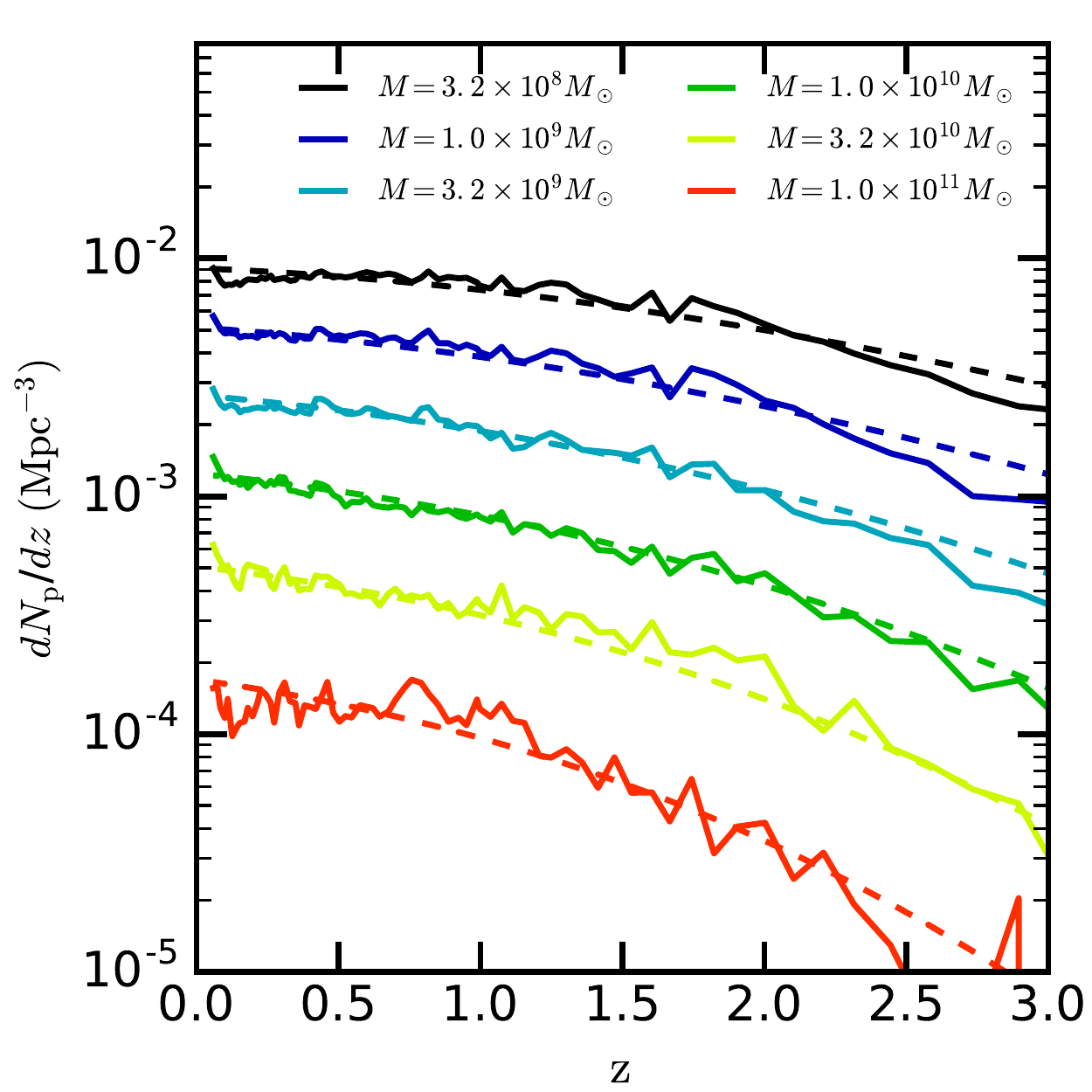}}}}
\caption{The promotion (left) and demotion (right) forward scatter rates.  
Solid lines indicate the promotion/demotion scatter rates as directly tabulated from the simulation; 
As in Figure~\ref{fig:destructive_merger_rate}, the promotion/demotion rates are shown for the same fixed $M_*$ at all redshifts, as indicated in the legends.
Dashed lines indicate the best fits of the form used in Appendix~\ref{sec:Fits} to assess the relative importance of galaxy coagulation and scattered growth rates. 
}
\label{fig:foward_scatter_rate}
\end{figure*}

\subsection{Scatter Rate}
\label{sec:scatter_rate}

Variable/stochastic growth rates of galaxies can drive mass rank order changes among a galaxy population.
In the ideal picture of constant comoving number density evolution all galaxies grow in mass with time at a rate that maintains mass rank ordering (i.e. no galaxies are allowed to ``pass" similar-mass systems in their mass rank order).
This assumption breaks down for real galaxy populations which experience stochastically varying growth rates owing to, e.g., the spread in the star formation main sequence and/or stellar mass growth via galaxy mergers.
Real galaxy populations will change mass rank when growth rates of similar mass galaxies are not homogeneous -- driven by either scatter in the star formation main sequence or ex-situ growth via stochastic galaxy mergers.

Unlike galaxy coagulation, the sense of mass rank order change owing to scattered growth rates can be either positive or negative.  
If a tracked galaxy population passes a large number of galaxies in mass rank order, 
fewer systems will remain with higher masses which results in a number density evolution rate of the same sign as galaxy mergers (positive\footnote{A tracked galaxy population growing rapidly with time will move forward {\it smaller} number density values with time.  However,  the sign of the rate of change is positive since Equation~\ref{eqn:dn_dz} expresses the number density evolution rate with redshift.}, according to Equation~\ref{eqn:dn_dz}).
Conversely, if a tracked galaxy population is passed by more systems, the number of systems with larger masses will grow with time leading to a number density evolution rate of the opposite sign as galaxy mergers.

To handle these two cases, we identify the median mass of a tracked galaxy population as being $\left< M_i \right>$ at time $t$ and $\left< M_i^\prime \right>$ at some later time $t^\prime$.
We can then define
\begin{itemize} 
 \item[1)] {\bf Galaxy mass rank demotion}: the number of galaxies with mass $M< \left< M_{i} \right> $ at time $t$ that have mass $M^\prime > \left< M_i^\prime \right>$ at $t^\prime$.
 \item[2)] {\bf Galaxy mass rank promotion}: the number of galaxies with mass $M> \left< M_{i} \right> $ at time $t$ that have mass $M^\prime < \left< M_i^\prime\right> $ at $t^\prime$.
\end{itemize}
We adopt the terms ``galaxy demotion" and ``galaxy promotion" since the tracked galaxy population is being demoted or promoted in mass rank ordering relative to the rest of the galaxy population.
We tabulate the galaxy promotion/demotion rates in the simulation using finite differencing with spacing of $\Delta z \approx 0.1$.  
In practice, we find all galaxies with $M>10^8 M_\odot$ at redshift $z$ and use the merger tree to identify all descendants at redshift $z-\Delta z$.
The galaxy promotion rate has the same sign as the galaxy merger rate (positive), while the galaxy demotion rate has the opposite.
Note that both in situ mass growth as well as growth via mergers are important to the scatter rate. 
However, galaxies that are not the main progenitor at time $t$ of their descendent at time $t^\prime$ are considered to have been ``consumed", and therefore do not contribute to the scatter rate.
The forward promotion and demotion mass rank scatter rates are shown in the left and right panels of Figure~\ref{fig:foward_scatter_rate} as solid lines for several mass bins.

The relative values of the galaxy promotion and demotion rates are set by the average mass growth rates of the tracked galaxy populations, the average mass growth rates of galaxies of neighboring mass (number density) bins, as well as the relative abundance of galaxies in the tracked and neighboring mass growth bins.
At redshift $z=0$, the galaxy promotion rate is larger than the galaxy demotion rate by a factor of $\sim 2$ for $M\approx3\times10^8 \msun$ systems but smaller by factor of $\sim 1.5$ for $M\approx 10^{11} \msun$ systems.
The point of equality (i.e. no net change from scatter in the median number density of the tracked galaxy population) between the promotion and demotion rates is around $M=3\times10^{10} \msun$.

We explore the relative importance of the galaxy promotion, demotion, and coagulation rates in more depth in the following subsection.

\subsection{Relative importance of galaxy coagulation and scattered growth rates}
\label{sec:RelativeRoles}

To compare the relative importance of the galaxy coagulation and scattered growth rates, we adopt simple polynomial fits that are shown as dashed lines in Figures~\ref{fig:destructive_merger_rate} and~\ref{fig:foward_scatter_rate}.
The form of the fits take the same fitting formula as the scatter and survival fraction fits discussed in Appendix~\ref{sec:Fits}.

The top panel of Figure~\ref{fig:compare_relative_importance} shows the net number density rate of change for galaxies for four mass bins (as indicated in the legend).
Most galaxies experience positive number density evolution rates which is driven by the combined dominance of mergers and mass rank promotion -- both of which drive galaxies to lower number densities as they move to lower redshift.
However, there are periods at early times where mass rank demotion dominates for massive galaxies (i.e. lower mass galaxies are passing this bin) leading to a galaxy population moving to lower mass rank (higher number density) with time.
These periods of time are indicated with dashed lines in Figure~\ref{fig:compare_relative_importance}.

\begin{figure}
\centerline{\vbox{\hbox{
\includegraphics[width=3.25in]{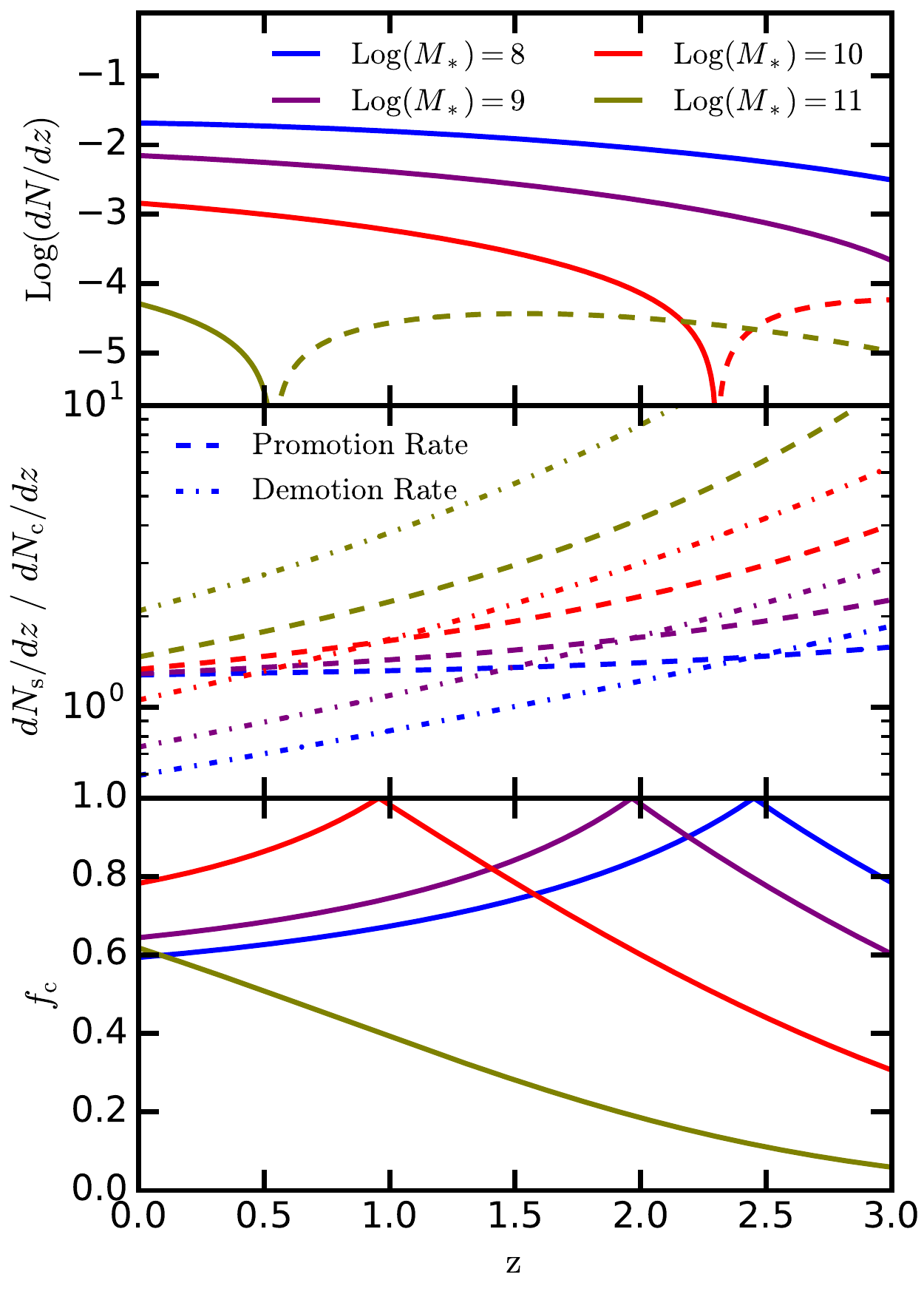}}}}
\caption{
(top) The net median rate of change of four galaxy populations of fixed stellar mass with time.  
Dashed lines indicate negative values -- where the galaxy demotion rate dominates.
Negative values only occur for galaxies which are at the upper end of the (redshift dependent) mass function.
(middle) The magnitude of the galaxy promotion and demotion rate relative to the galaxy coagulation rate.
(bottom) The fractional importance of galaxy coagulation to the net median number density evolution rate.
  }
\label{fig:compare_relative_importance}
\end{figure}

The net number density evolution rate can drop to zero and/or change signs depending on the relative importance of the galaxy coagulation, promotion, and demotion rates.
We therefore address the relative contribution of mergers and scatter by considering the magnitude of the promotion and demotion rates (each shown individually) normalized by the galaxy coagulation rate as shown in the middle panel of Figure~\ref{fig:compare_relative_importance}.
With the exception of the demotion rates of the two lowest mass bins we find that the promotion- and demotion-scatter rates tend to be larger than the galaxy coagulation rate.   
The galaxy scatter rates are an order of magnitude larger than the coagulation rates for high mass galaxies at high redshift.  
However, since the promotion and demotion rates partially cancel it is possible for the coagulation rates to still dominate the net median mass rank order evolution.

The bottom panel of Figure~\ref{fig:compare_relative_importance} shows the fractional contribution of mergers to the total number density evolution rate. 
Here, we have specifically defined the merger contribution to the total number density evolution rate as
\begin{equation}
f_c = \frac{ \left< dN_c / dz \right> }{\left| \left< dN_c / dz \right> \right|  + \left| \left< dN_s / dz \right> \right|   }.
\end{equation}
Galaxy coagulation is the dominant contributor to the net number density evolution rate for galaxies with stellar masses $M_* < 10^{9.5} \msun$ out to redshift $z=3$.
For galaxies that are more massive than $10^{9.5} \msun$, the galaxy coagulation will be subdominant to scatter at early times, but eventually comes to dominate.  
However, for the most massive systems, scatter naturally dominates the mass rank order evolution budget owing to 
(i) the paucity of more massive systems that are able to drive large galaxy coagulation rates and 
(ii) the dominance of ex-situ stellar mass growth driven by stochastic merger events with lower mass galaxies for massive, quenched systems.

Massive galaxies in the Illustris simulation grow predominantly through galaxy mergers~\citep[][]{RodriguezGomez2015}.
Yet, scatter in their rank order dominates over galaxy coagulation when considering their mass rank order evolution budget.
These two claims are consistent because the ex-situ growth of massive galaxies is driven by mergers with less massive systems.
Mergers with less massive systems -- even if they are abundant and significantly impact the galaxy's mass evolution -- do not influence the galaxy coagulation rate.
Instead, the stochastic growth histories of massive galaxies associated with galaxy mergers only acts to increase the influence of mass rank order scatter on the total mass rank order evolution budget.

A converse argument is true for low mass galaxies which build up the majority of their mass through in-situ star formation while their total mass rank order evolution budget is dominated by galaxy coagulation.
Specifically, the galaxy mergers that lead to the large galaxy coagulation rate for low mass galaxies do not involve the low mass galaxies themselves.
Instead, the galaxy coagulation rate is set by the merger rate between two larger galaxies and the galaxy scatter rate is set by the scatter in the in-situ mass growth rates.

\section{Discussion}
\label{sec:Discussion}
Forging progenitor/descendant links between observed galaxy populations at different redshifts using a constant cumulative number density is widely used and reasonably physically justified.
It has been shown using abundance matching models, semi-analytic models, and hydrodynamical simulations that one can recover, e.g., the mass evolution of Milky Way progenitors while only introducing errors of order $\sim0.3-0.5$ dex.
However, the sense of this error is systematic and partially correctable.
In~\citet{Torrey2015b} and this paper, we have argued for a modified method of forging progenitor/descendant links using relaxed assumptions about the evolution of galaxies in comoving number density space based on the results of numerical simulations.
We have advocated that progenitor/descendant galaxy populations can be statistically linked based on their comoving number density.
However, we also argue that this link must include both the median evolution of the galaxy population in comoving number density space~\citep{Torrey2015b} and the significant intrinsic scatter (this paper).

In this paper we analyzed galaxy evolution in number density space using basic analytic arguments and empirical fits derived from the Illustris simulation.
The empirical fits include a direct measure of the intrinsic scatter that makes identifying {\it direct} progenitor/descendant links impossible.
Accounting for the intrinsic scatter makes it possible to draw conclusions about the statistical distribution of possible progenitor/descendant properties that would not be captured using a single comoving number density track.
This is critical for describing the full range of formation histories that galaxies may evolve through.
Recognizing the significant impact of scattered growth rates on the formation of galaxy populations is important for avoiding a secondary form of progenitor bias:  considering only the median growth track to the exclusion of all others.

We have made an attempt in this paper to reconcile the asymmetry that is encountered when tracking galaxy populations forward and backward in time.
It is well-established that tracking galaxies forward and backward in time yields median number density evolution tracks that are not symmetric.
We have shown that using the single DDF prescribed in Section~\ref{sec:DDF}, one can broadly recover forward evolution of galaxies in number density space, and approximately recover the backward/progenitor distribution function.
This is important because it allows us to view the number density evolution of galaxies as a single coherent process, regardless of whether one is tracking galaxies forward or backward in time.

\subsection{Dependence on Baryon Physics}
The fits presented in Figure~\ref{fig:foward_num_dens_evo} are very similar to those obtained by~\citet{Behroozi2013} which were obtained using abundance matching on a dark matter only simulation~\citep{Torrey2015b}.
This indicates that our fits are primarily driven by the stochastic assembly of dark matter underlying dark matter haloes, 
rather than the specific physics implementation included in the Illustris simulations.
This conclusion is also supported by the similar growth tracks that have been reported in semi-analytic models~\citep[e.g.,][]{Leja2013, Terrazas2016} and other hydro simulations~\citep[e.g.,][]{Jaacks2015}.
However, there is value in confirming and updating the specific fits presented in this paper using other simulation methods involving different assumptions about the baryonic physics and within other (particularly, larger) simulation volumes.

\subsection{Application to observational datasets}
The evolution tracks prescribed in this paper can be used to link selected galaxy populations to progenitor or descendant galaxy {\it populations}.  
In contrast to linking at a constant comoving number density, or at a non-constant comoving number density~\citep{Torrey2015b}, the PDFs and DDFs will identify a wider range of possible galaxy progenitors/descendants.
The spread in the progenitor/descendant population properties can be set based on results of numerical simulations.

A detailed exploration of how our prescriptions can be applied to observational data is left to a companion paper~\citep{Wellons2016}.
There, we describe in detail and validate a straightforward method to translate the DDFs tabulated in this paper into mass, size, and star formation rate evolution tracks based directly on observational data.

\section{Conclusions}
\label{sec:Conclusions}
In this paper we have critically examined the nature of galaxy evolution in comoving number density space.
We have provided a detailed description of how the median number density and standard deviation for a tracked galaxy population evolves with time.
The primary conclusions of this paper are as follows:
\begin{enumerate}
\item[1)] We defined the descendant distribution function (DDF) and progenitor distribution function (PDF) to capture the median number density evolution, the spread among the tracked galaxy population, and the survival fraction of tracked galaxies.  
Practical fits to the DDF and PDF were provided which can be applied to observational data. 
Publicly available {\small PYTHON} scripts to evaluate the fitting functions have been provided.\footnote{\url{https://github.com/ptorrey/torrey_cmf}}
\item[2)] The median number density evolution described in the DDF and PDF are not symmetric.
This asymmetry is driven by the convolution of the DDF with the relative abundance of progenitor galaxies that is required to recover the PDF.  
A quantitative description of this is given in Equation~\ref{eqn:prog_avg_rank} which is demonstrated in Figure~\ref{fig:backward_num_dens_evo}.
\item[3)] The total DDF evolution rate can be broken down into a contribution from galaxy coagulation and galaxy scattered growth rates.  
We quantified the relative contribution of mergers and scatter, and found that galaxy coagulation is the dominant driver of the number density evolution rate for a range of mass bins and redshift ranges (described in Section~\ref{sec:RelativeRoles} and Figure~\ref{fig:compare_relative_importance}).
\item[4)] The evolution tracks for galaxy populations in number density space defined in this paper for redshift $z\leq3$ in the mass ranges $10^8 \msun < M_* $ can be used to analyze observational data sets.  
A forthcoming paper~\citep{Wellons2016} will practically demonstrate how the diversity of evolution tracks in number density space impact conclusions about the mass, star formation rate, and velocity dispersion evolution of galaxy populations.
\end{enumerate}

\section*{Acknowledgements}
Support for PFH was provided by an Alfred P. Sloan Research Fellowship, NASA ATP Grant NNX14AH35G, NSF Collaborative Research Grant \#1411920, and CAREER grant \#1455342. 
MV acknowledges support through an MIT RSC award.
The Illustris simulation was run on the CURIE supercomputer at CEA/France as part of PRACE project RA0844, and the SuperMUC computer at the Leibniz Computing Centre, Germany, as part of project pr85je. Further simulations were run on the Harvard Odyssey and CfA/ITC clusters, the Ranger and Stampede supercomputers at the Texas Advanced Computing Center through XSEDE, and the Kraken supercomputer at Oak Ridge National Laboratory through XSEDE. 
The analysis reported in this paper was performed on the Caltech compute cluster ``Zwicky" (NSF MRI award \#PHY-0960291), 
the joint partition of the MIT-Harvard computing cluster ``Odyssey" supported by MKI and FAS, 
and allocation TG-AST150059 granted by the Extreme Science and Engineering Discovery Environment (XSEDE) supported by the NSF.


\appendix

\section{Median Number Density, Scatter, and Survival Fraction Fitting Functions}
\label{sec:Fits}
In this Appendix we provide fits to the median number density, one-sigma scatter in the number density, and survival fraction of a tracked galaxy population based on the Illustris simulation.
The fits are not intended to carry physical meaning, but instead to serve as tools that can be used to link galaxy populations based on observational data sets.
A single fit for the forward evolution of galaxy populations is required for the evaluation of Equation~\ref{eqn:prog_avg_rank}.
The current fits have a large number of parameters because reduced parameters led to noticeable loss of fit accuracy.

In addition to publishing the fits here, {\small PYTHON} routines to evaluate these fits are provided through \url{https://github.com/ptorrey/torrey_cmf}.

\subsection{Forward Median Number Density Fit}
\label{sec:forward_med_fits}
The forward-tracked median number density evolution is well-fit with 
\begin{multline}
\left< \mathcal N(z) \right>  = \mathcal N_0 + \\ 
    \Delta z  \left( A_0 + A_1 \mathcal N_0 + A_2 \mathcal N_0 ^2 + A_3 \mathcal N_0 ^3   + A_4 \mathcal N_0 ^4 \right)  + \\ 
    \Delta z ^2 \left( B_0 + B_1 \mathcal N_0 + B_2 \mathcal N_0 ^2  + B_3 \mathcal N_0 ^3   + B_4 \mathcal N_0 ^4 \right) + \\
    \Delta z ^3 \left( C_0 + C_1 \mathcal N_0 + C_2 \mathcal N_0 ^2  + C_3 \mathcal N_0 ^3   + C_4 \mathcal N_0 ^4 \right) 
\label{eqn:foward_num_dens_evo}
\end{multline}
where $\mathcal N_0 = \mathrm{Log}(N(z_0))$ and $\Delta z =\left| z_0-z \right|$.  
The coefficients $A_i$, $B_i$, and $C_i$ are each functions of the initial selection redshift for the galaxy population according to 
$\alpha_i = \alpha_{i,0} + z_0 \alpha_{i,1} + z_0^2 \alpha_{i,2}$ (e.g., $A_1 = A_{1,0} + z_0 A_{1,1} + z_0^2 A_{1,2}$).
This results in a fitting function that accounts for the dependence of initial selection redshift, initial selection number density, and evolution time.
The fit has 45 free variables which are determined via linear regression based on the number density evolution of all galaxies with stellar masses greater than $10^8 M_\odot$ tracked forward in time from any initial selection redshift below $z\leq3$.  
This high number of coefficients is a result of making a single fit to simultaneous variations in $\langle N_f \rangle$, $N_0$, $z_0$, and $\Delta z$.
The best fit coefficients are given in Table~\ref{table:foward_num_dens_evo}, the resulting fits are shown as dashed black lines in Figure~\ref{fig:foward_num_dens_evo}, and easily applied Python functions to evaluate these fits are publicly available.\footnote{\url{https://github.com/ptorrey/torrey_cmf}}
Equation~\ref{eqn:foward_num_dens_evo} describes the median comoving number density evolution for galaxy population selected at any redshift $z\leq3$ in the mass ranges $10^8 \msun < M_* $ tracked forward in time.
This single fit accurately captures the number density evolution for a wide range of galaxies owing to the high order of the polynominal expansion.

\begin{table}
\begin{center}
\caption{ Best-fit parameters to the forward median number density evolution (Eqn~\ref{eqn:foward_num_dens_evo}), scatter number density evolution (Eqn~\ref{eqn:foward_scatter_evo}), and galaxy survival fraction (Eqn~\ref{eqn:foward_surv_evo}).  
The resulting best fits are demonstrated in Figure~\ref{fig:foward_num_dens_evo}.}
\label{table:foward_num_dens_evo}
\begin{tabular}{ c  c c c c c  }
\hline
  &      \multicolumn{1}{|c|}{$j=0$}     &       \multicolumn{1}{|c|}{$j=1$}      &     \multicolumn{1}{|c|}{$j=2$}     \\
\hline 
\hline 
$A_{0,j}$      & -6.61692  & 4.77210 &  -0.62078   \\
$A_{1,j}$      & -12.55386  & 10.06238 &  -1.59352   \\
$A_{2,j}$      & -8.49063  & 7.30574 &  -1.29725 \\
$A_{3,j}$      & -2.39311  & 2.17088 &  -0.41597  \\
$A_{4,j}$      & -0.23725  & 0.22454 &  -0.04541   \\
$B_{0,j}$      & 20.21598  & -19.23783 &  3.78653  \\
$B_{1,j}$      & 35.82274  & -34.93341 &  7.11385 \\
$B_{2,j}$      & 22.31793  & -22.30232 &  4.67740 \\
$B_{3,j}$      & 5.84523  & -5.96409 &  1.28203  \\
$B_{4,j}$      & 0.54639  & -0.56617 &  0.12411  \\
$C_{0,j}$      & -12.51657  & 11.75015 &  -2.43751  \\
$C_{1,j}$      & -21.96728  & 20.80609 &  -4.36787  \\
$C_{2,j}$      & -13.61242  & 13.01445 &  -2.76350  \\
$C_{3,j}$      & -3.56242  & 3.43187 &  -0.73592  \\
$C_{4,j}$      & -0.33457  & 0.32361 &  -0.06990  \\
\hline
$D_{0,j}$      & 0.30754  & -0.35706 &  0.07812 \\
$D_{1,j}$      & 0.11659  & -0.36310 &  0.08781 \\
$D_{2,j}$      & 0.05432  & -0.08568 &  0.01939 \\
$E_{0,j}$      & 0.34105  & -0.26566 &  0.05867 \\
$E_{1,j}$      & 0.40405  & -0.27861 &  0.05734 \\
$E_{2,j}$      & 0.06319  & -0.04606 &  0.01022 \\
\hline
$F_{0,j}$      & -0.57610  & 0.25395 &  -0.06909 \\
$F_{1,j}$      & -0.32307  & 0.30592 &  -0.08416 \\
$F_{2,j}$      & -0.05880  & 0.07159 &  -0.01936 \\
$G_{0,j}$      & 0.34438  & -0.21727 &  0.04129 \\
$G_{1,j}$      & 0.29859  & -0.21186 &  0.04228 \\
$G_{2,j}$      & 0.03889  & -0.03177 &  0.00690 \\
\hline
\hline 
\end{tabular}
\end{center}
\end{table}

\subsection{Forward Number Density Scatter Fit}
\label{sec:forward_scat_fits}
We quantify the scatter evolution in the DDF using the standard deviation of the number density distribution for the tracked galaxy population in log space.
The fitting function
\begin{multline}
\sigma(z)  = \sigma_0 + \Delta z  \left( D_0 + D_1 \mathcal N_0 + D_2 \mathcal N_0 ^2  \right)  + \\ \Delta z ^2 \left( E_0 + E_1 \mathcal N_0 + E_2 \mathcal N_0 ^2  \right)
\label{eqn:foward_scatter_evo}
\end{multline}
describes the forward-tracked lognormal standard deviation well.
The coefficients $D_i$ and $E_i$ are functions of the selection redshift of the galaxy population, $\alpha_i = \alpha_{i,0} + z_0 \alpha_{i,1} + z_0^2 \alpha_{i,2}$,
and are set via a linear regression from the same Illustris galaxies used in Appendix~\ref{sec:forward_med_fits}.
The best fit values for the $D_i$ and $E_i$ coefficients are given in Table~\ref{table:foward_num_dens_evo}.

\subsection{Forward Survival Fraction Fit}
\label{sec:forward_surv_fits}
We quantify the survival fraction of tracked galaxy populations by calculating the fraction of systems which are consumed by a larger galaxy.
The fitting function
\begin{multline}
f_{\mathrm{s}}(z)  = 1.0 + \Delta z  \left( F_0 + F_1 \mathcal N_0 + F_2 \mathcal N_0 ^2  \right)  + \\ \Delta z ^2 \left( G_0 + G_1 \mathcal N_0 + G_2 \mathcal N_0 ^2  \right)
\label{eqn:foward_surv_evo}
\end{multline}
describes the forward tracked survival fraction well.
The best fit values for the $F_i$ and $G_i$ coefficients are given in Table~\ref{table:foward_num_dens_evo}.

\subsection{Backward Median Number Density Fit}
We showed in section~\ref{sec:PDF} that the PDF can be broadly recovered using the DDF and forward scatter fits.
However, we also argued that the accuracy of this recovery procedure is limited by deviations of the DDF from a strictly log-normal distribution, and by progenitors drawn from outside the valid fitting regions (as set by the resolution and box-size of Illustris).
For consistency with the fits presented in the previous subsection we also fit to the median number density for the progenitor distribution function using
\begin{multline}
 \left< \mathcal N(z) \right>^\prime  = \mathcal N_0 + \Delta z  \left( A_0^\prime + A_1^\prime \mathcal N_0 + A_2^\prime \mathcal N_0 ^2  \right)  + \\ 
                        \Delta z ^2 \left( B_0 ^\prime+ B_1^\prime \mathcal N_0 + B_2^\prime \mathcal N_0 ^2  \right)
\label{eqn:backward_num_dens_evo}
\end{multline}
where as before $\mathcal N_0 = \mathrm{Log}(N(z_0))$ and $\Delta z =\left| z_0-z \right|$.  
Primes in Equation~\ref{eqn:backward_num_dens_evo} are used to denote fits to the progenitor distribution.
We assume galaxy populations are selected at redshift $z=0$ and therefore we can directly specify with only 6 parameters.
Best fit values for Equation~\ref{eqn:backward_num_dens_evo} are given in Table~\ref{table:backward_num_dens_evo} and the resulting best fit tracks are demonstrated with black dashed lines in Figure~\ref{fig:backward_num_dens_evo}.

\subsection{Backward Number Density Scatter Fit}
We fit to the one sigma standard deviation in the number density evolution of backward tracked galaxy populations using
\begin{multline}
\sigma(z)  = \sigma_0 + \Delta z  \left( D_0^\prime + D_1^\prime \mathcal N_0 + D_2^\prime \mathcal N_0 ^2  \right)  + \\ \Delta z ^2 \left( E_0^\prime + E_1^\prime \mathcal N_0 + E_2^\prime \mathcal N_0 ^2  \right).
\label{eqn:backward_scatter_evo}
\end{multline}
Best fit values for Equation~\ref{eqn:backward_num_dens_evo} are given in Table~\ref{table:backward_num_dens_evo} and the resulting best fit tracks are demonstrated with black dashed lines in Figure~\ref{fig:backward_num_dens_evo}.

\begin{table}
\begin{center}
\caption{ Best-fit parameters to the backward median number density evolution.  
The resulting best fits are demonstrated in Figure~\ref{fig:backward_num_dens_evo}.}
\label{table:backward_num_dens_evo}
\begin{tabular}{ c  c c c c c  }
\hline
  &      \multicolumn{1}{|c|}{$j=0$}     &       \multicolumn{1}{|c|}{$j=1$}      &     \multicolumn{1}{|c|}{$j=2$}     \\
\hline 
\hline 
$A_j^\prime$ & -0.392746  &  -0.473089  &  -0.097202 \\ 
$B_j^\prime$ & 0.188759  &  0.165314  &  0.034512 \\ 
\hline
$D_j^\prime$ & 0.009616  &  -0.068805  &  0.004060 \\ 
$E_j^\prime$ & -0.019706  &  -0.011360  &  -0.005398 \\ 
\hline
\hline 
\end{tabular}
\end{center}
\end{table}

\end{document}